\documentclass[aps,prl,twocolumn,superscriptaddress,showpacs]{revtex4}
\usepackage{graphics}

\bibliography{20}

\begin{document}

\title{Localized Modes in Open One-Dimensional Dissipative Random Systems}
\author{K. Yu. Bliokh}
\affiliation{Institute of Radio Astronomy, 4 Krasnoznamyonnaya st., Kharkov, 61002,
Ukraine}
\affiliation{Department of Physics, Bar-Ilan University, Ramat-Gan, 52900, Israel}
\author{Yu. P. Bliokh}
\affiliation{Physics Department, Technion-Israel Institute of Technology, Haifa, 32000,
Israel}
\author{V. Freilikher}
\affiliation{Department of Physics, Bar-Ilan University, Ramat-Gan, 52900, Israel}
\author{A. Z. Genack}
\affiliation{Department of Physics, Queens College of the City University of New York,
Flushing, New York 11367l}
\author{B. Hu}
\affiliation{Department of Physics, Queens College of the City University of New York,
Flushing, New York 11367l}
\author{P. Sebbah}
\affiliation{Laboratoire de Physique de la Mati{\`{e}}re Condens{\'{e}}e, CNRS UMR6622
and Universit{\'{e}} de Nice - Sophia Antipolis, Parc Valrose, 06108, Nice
Cedex 02, France}

\begin{abstract}
We consider, both theoretically and experimentally, the excitation and
detection of the localized quasi-modes (resonances) in an open dissipative
1D random system. We show that even though the amplitude of transmission
drops dramatically so that it cannot be observed in the presence of small
losses, resonances are still clearly exhibited in reflection. Surprisingly,
small losses essentially improve conditions for the detection of resonances
in reflection as compared with the lossless case. An algorithm is proposed
and tested to retrieve sample parameters and resonances characteristics
inside the random system exclusively from reflection measurements.
\end{abstract}

\pacs{42.25.Dd, 78.70.Gq, 78.90.+t}
\maketitle

Random stratified media are found in numerous geological and biological
settings as well as in fabricated materials. Wave interactions in such
systems determine, for example, reflection and transmission from multilayer
dielectric stacks used as optical reflectors, filters and lasers, and
propagation of seismic waves in the earth's crust, microwave radiation in
sand layers, and sonic waves in the oceans. In addition, though considerable
effort has been expended to develop highly periodic structures, deviations
from periodicity can significantly modify the characteristics of optical and
microwave tunable photonic crystals. At the same time, it may be possible to
utilize highly disordered samples, for many applications. For example,
tunable switches or narrow line laser sources \cite{Milner} can be created
in randomly stacked systems or in disordered fibers. Since the location,
width, and intensity of resonance are random, these can be found only by
direct measurement of the field inside the sample, which is generally not
feasible. Although strong localization of waves and resonances in
one-dimensional random media has been extensively studied theoretically \cite%
{Lifshits, Sheng, Frish}, most of the analytical results were obtained for
lossless systems and for values averaged over ensembles of random
realizations. But results for an ensemble do not provide the intensity
spectrum within a particular sample, which is often essential.

In this letter, we consider samples, in which dissipation is weak in the
sence that the condition 
\begin{equation}
{\frac{\Gamma _{t}}{v_{g}}}l_{loc}\equiv \Gamma l_{loc}\ll \Gamma L\ll 1,
\label{1}
\end{equation}%
is satisfied. Here, $\Gamma _{t}$ and $\Gamma $, are, respectively, the
temporal and spatial decrements of the wave energy due to loss, $v_{g}$ is
the local group velocity of the wave, $L$ is the sample length, and $l_{loc}$
is the localization length.

The nature of wave propagation in absorbing random systems is
illustrated by measurements in a single-mode microwave rectangular waveguide
filled with random elements. Samples are composed of random mixtures of 31
randomly oriented elements which are 8 mm in length and are comprised of a
solid ceramic first half and a second half which is milled to create an air
space between two thin ceramic walls of thickness 0.8 mm, five 4-mm-long,
high dielectric material slabs, identical to the first half of the binary
element, and five 4 mm-long, low dielectric Styrofoam slabs. Only with the
introduction of the 4 mm elements are states introduced deep within the band
gap.The samples are prepared by choosing a random order of the elements and
a random orientation of the binary element (ceramic part facing eather input
or output) in the waveguide. Comparison of measurements of the frequencies
of the first mode at the two band edges of the band gap in a periodic
structure of binary elements with a one dimensional scattering matrix
simulation which includes waveguide dispersion gives n1 $\sim$1.67 and
n2 $\sim$1.08 for the refractive indices of the first and second
portions of these elements, respectively. The field inside the waveguide is
detected along a 2 mm wide slot along the length of the top of the
waveguide. The field is weakly coupled to an antenna so that it is not
affected by the measurement. The core of the coaxial adapter without a
protruding antenna picks up the field inside an enclosure coupled to the
waveguide through a 2 mm hole which is pressed against the slot. For each
random configuration, the detector is translated in 1 mm steps along the
slot. The in and out of phase components of the field were collected at each
position with use of an HP8720C vector network analyzer. Spectra over 6 GHz
were collected in 3.5 MHz steps. The cutoff frequency of the empty waveguide
is 13.5 GHz. 
The frequency range covered is slightly larger than the stop
band of the periodic structure. 

\begin{figure}[tbh]
\centering \scalebox{0.40}{\includegraphics{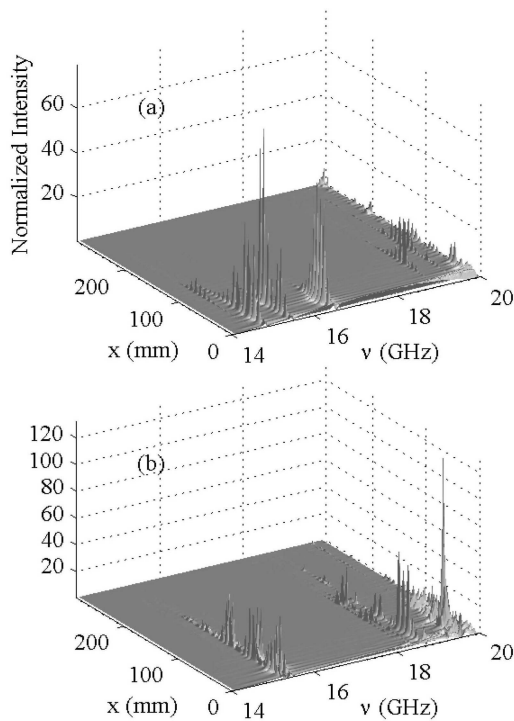}}
\caption{Intensity VS frequency and position inside the sample for two
different samples.}
\label{Fig1}
\end{figure}

Figure 1 shows the intensity, $I(x,\nu)$, inside two random configurations
normalized to the intensity of the incident wave as a function of coordinate
and frequency. Distinct localized states with high intensity are seen in
Fig.~1a. In addition, multiple peaks in space such as those shown in Fig. 1b
are found in other configurations. This field distribution has been
discussed theoretically by Lifshits and Kirpichenkov and
by Pendry \cite{Lifshits2}. Such a distribution has been termed a ``necklace
state'', though they correspond to several spectrally overlapping states
rather than to a single state \cite{Sebbah}. These states are associated
with non-Lorentzian spectra, which have been observed in optical
transmission through random layered media \cite{Sebbah, Bertolotti}.

The reflectance spectrum, $R(\nu ),$ for the distributions Fig. 1a is shown
in Fig. 2a. The instrumental error of the measurements of the reflection
coefficient was about $5\%.$ The transmittance (intensity at the output), $%
T(\nu )\equiv I(L,\nu )$, within a spectral range which encompasses only
exponentially peaks modes on resonance and evanescent waves off resonance is
below the experimental noise level (Fig.~2b). Sharp minima are observed in $%
R(\nu )$ that are not accompanied by an essential increase of $T(\nu )$
(note the difference in the scales in Figs.~2a and 2b). This is in contrast
to the lossless case where $R$ and $T$ are connected by energy conservation, 
$R+T=1$. 
\begin{figure}[tbh]
\centering \scalebox{0.44}{\includegraphics{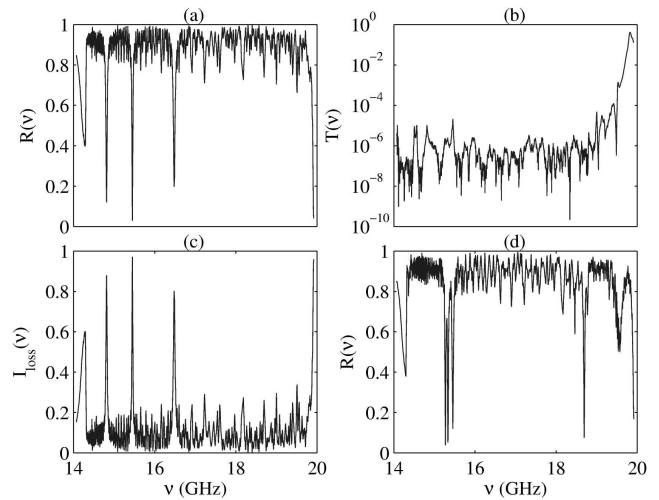}}
\caption{Spectra of (a) reflection, (b) transmission, and (c) normalized
energy loss, $I_{loss}(\protect\nu )$, for Fig.~1a. The reflection spectrum
of the sample with the distribution shown in Fig.~1b is shown in (d).}
\label{Fig2}
\end{figure}
Dips in $R(\nu )$ arise because the energy lost at each point within the
sample is proportional to the product of the intensity of the field at that
point and spatial absorption coefficient. Energy conservation requires, 
\begin{equation}
1-T-R=I_{loss}=\Gamma \int_{0}^{L}{\frac{v_{g}}{v_{g0}}}I\left( x,\nu
\right) \varepsilon (x)dx,  \nonumber  \label{2}
\end{equation}%
where $I_{loss}$ is the rate of energy dissipated in the sample, $%
\varepsilon (x)$ is the dielectric constant, and $v_{g0}$ is the group
velocity in the empty waveguide. For the following estimations, permittivity 
$\varepsilon (x)$ is replaced by its mean value $\bar{\varepsilon}$, whereas 
$v_{g0}\simeq c$ and $v_{g}\simeq c/\sqrt{\bar{\varepsilon}}$. Then, the
rate of energy dissipation can be written as $I_{loss}\simeq \Gamma \sqrt{%
\bar{\varepsilon}}\int_{0}^{L}I\left( x,\nu \right) dx$.

Off resonance, the ensemble average intensity falls exponentially with decay
length $l_{loc}$. The energy loss can therefore be estimated as $%
I_{loss}\sim \Gamma l_{loc}\sqrt{\bar{\varepsilon}}\ll 1$. Since the
off-resonance transmittance $T\sim T_{typ}\equiv \exp \left(
-L/l_{loc}\right) \ll 1$, the reflectivity $R $ is close to unity, as it
would be in the lossless case, but it saturates at $R\sim R_{typ}=1-\Gamma
l_{loc}\sqrt{\bar{\varepsilon}}-T_{typ}$.

On resonance, loss can be of order unity due to the high intensity in
localized states, even when Eq.~(\ref{1}) holds (Fig.~2c). This suppresses
both the reflected and transmitted fluxes. In Fig. 2, weak loss is seen to
suppress the transmitted wave on resonance below experimental noise, while
the suppression of the reflected wave is easily observed. The ``necklace''
shown in Fig.~1b is also easily detected in reflection as a group of closely
spaced peaks with the number of peaks equal to that of the coupled states
underlying the ``necklace'' (Fig.~2d).

To interpret the experimental data presented above, we extended the approach
developed in \cite{Bliokh} for modelling lossless random systems by
incorporating small absorption in the model. The approach is based on
associating each localized state at a frequency $\nu $ with a structure
comprised of an essentially transparent segment (``well'') of small length $%
l\ll L$, surrounded by essentially non-transparent segments (``barriers'')
with small transmittance $T_{1,2}\ll 1$ characteristic of ``typical''
non-resonant configurations. At the frequency $\nu $ the configuration forms
a high-Q resonator and wave propagation through it can be treated as a
particular case of the general problem of the transmission through an open
resonant system, regular or random, whether it is quantum-mechanical
potential well, optical or microwave resonator or 1D random medium. The
distinguishing feature of the random medium is that there are no regular
walls, and the separate transmission coefficients of the isolated left and
right segments of the sample, $T_{1}$ and $T_{2}$, respectively, are
exponentially small as a result of Anderson localization.

A necklace state exists in a random configuration when at some frequency
there are several effective cavities separated by typical non-transparent
segments . Generalization of the theory to this case amounts to the
solution of a bit more complicated problem of the
resonant propagation through a set of spatially separated potential wells
(coupled resonators) whose number is equal to the number of dips in the
measured frequency dependence of the reflection coefficient (Fig. 1b). It is
beyond the scope of this letter and will be considered in a future
publication. 

Calculations of the resonant transmittance and reflectance of an open
resonator with the absorption rate $\Gamma $, and of the intensity inside
the effective cavity yield
\begin{equation}
T_{res}={\frac{4T_{1}T_{2}}{\left( \Gamma \mathit{l}\sqrt{\bar{\varepsilon}}%
+T_{1}+T_{2}\right) ^{2}}},  \label{3a}
\end{equation}%
\begin{equation}
R_{res}=1-T_{res}-\Gamma l\sqrt{\bar{\varepsilon}}T_{res}/T_{2},  \label{3b}
\end{equation}%
\begin{equation}
I_{res}=T_{res}/T_{2}.  \label{3c}
\end{equation}%
The half-width of the resonant peaks in $R\left( \nu \right) $ is given by 
\begin{equation}
\delta \nu =\frac{c}{4\pi l\bar{\varepsilon}}\left( \Gamma l\sqrt{\bar{%
\varepsilon}}+T_{1}+T_{2}\right) .  \label{4}
\end{equation}

In the case of Anderson localization in a disordered sample, coefficients $%
T_{1,2}$ are approximately given by $T_{1,2}\simeq\exp\left[-\left(L/2\pm
d\right)/l_{loc}\right]$, where $d$ is the coordinate of the resonance
relative to the center of the sample, so that $T_{1}T_{2}\simeq T_{typ}$.
Eqs.~(\ref{3a})--(\ref{3c}) satisfy the conservation law Eq.~(\ref{2}) if $%
I_{loss}\simeq I_{res}\Gamma l\sqrt{\bar{\varepsilon}}$. The reflection
coefficient, $R_{res}$, and the rescaled intensity at the resonance, $%
I_{res}^{\ast }=T_{typ}^{1/2}I_{res}$, depend on the position of the cavity
and on the absorption through universal functions of two dimensionless
parameters $a=\exp \left( -d/l_{loc}\right) $ and $b=2\Gamma l\sqrt{\bar{%
\varepsilon}}T_{typ}^{-1/2}$: 
\begin{equation}
R_{res}=\left( \frac{b+a^{-1}-a}{b+a^{-1}+a}\right) ^{2},~I_{res}^{\ast }=%
\frac{8a}{\left( b+a^{-1}+a\right) ^{2}}.  \label{6}
\end{equation}

Figs.~3a and 3b depict the dependencies of $I_{res}^{\ast }$ and $R_{res},$
on parameters $a$ and $b$. The resonances can be detected by peaks in the
reflection spectrum, $R(\nu )$. The bright area in Fig.~3a corresponds to
well-excited localized states with high values of the intensity $I_{res}$,
whereas the bright area in Fig.~3b indicates easily detectable resonances
with pronounced minima in $R(\nu )$. 

\begin{figure}[tbh]
\centering \frame{} \scalebox{0.42}{\includegraphics{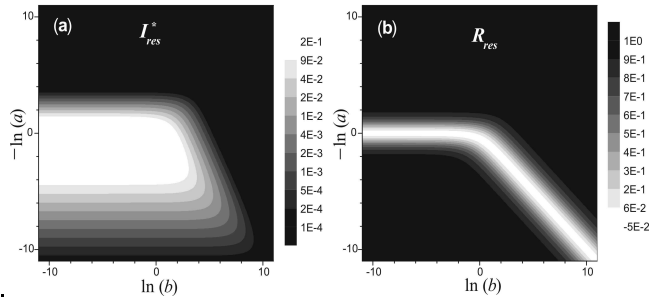}}
\caption{The rescaled resonant intensity (a) and reflection coefficient (b)
via parameters $a$ and $b$}
\label{Fig3}
\end{figure}

If dissipation is extremely weak so that $b\ll 1$, the influence of
dissipation on the resonances is negligible, $I_{loss}\ll 1$, and
strongly-excited modes are located near the center and in the first, input
half of the sample (Fig.~3a) as is in a lossless system \cite{Bliokh}. It
follows from Fig.~3b that for $b\ll 1$, easily-detected modes are excited
only in a relatively small region near the center of the sample. Thus, only
a relatively small part of excited resonances can be detected via $R(\nu )$
or $T(\nu )$ measurements.

If $b\gg 1$ [but Eq.~(\ref{1}) still holds], $I_{loss}\sim 1$ and small
losses strongly affect the resonance characteristics. The absorption
substantially reduces resonance transmission, $T_{res}\ll 1$, Eq.~(3), and
also gives rise to the non-monotonic intensity distribution inside the first
(close to the input) effective barrier. The intensity of the resonances
decreases with increasing $b$ (Fig.~3a), and the region, where well-excited
resonances are located, becomes narrower and closer to the input. At the
same time, the region in which resonant states can be detected becomes 
\textit{larger} and also shifts towards the input (Fig.~3b). For $b\gg 1$
almost all well-excited resonances are detectable via $R(\nu )$
measurements. Thus, the ``detectability'' of resonances is \textit{improved}
when small dissipation is present. It can be shown from Eqs.~(\ref{6}) that
the number of resonances that can be detected in reflection is twice as
large when $b\gg 1$ than in the opposite case ($b\ll 1$) for any given
minimum in the discernible value of $R$. Moreover, at different values of $b$
(or $\Gamma$) resonances in different regions of the sample are excited and
detectable, which can provide for the scan of the sample through variations
of losses. Thus, surprisingly, the presence of loss can make it easier to
find the characteristics of the eigenmodes of a random sample.

It is important to notice, that for $b\gg 1$ the reflection coefficient at
resonances is a non-monotonic function of dissipation (Fig.~3b), and reaches
zero when 
\begin{equation}
b=a-a^{-1}.  \label{CC}
\end{equation}
This effect is well-known in optics and microwave electronics as critical
coupling \cite{Yariv}. Since no energy is reflected from the sample at
critical coupling, it corresponds to the resonance with the highest
intensity $I_{res}$. Eq.~(\ref{CC}) determines
the position, $x_{c}=L/2+d$, of the most powerful resonance in a sample with
a given $\Gamma $: 
\begin{equation}  \label{5}
x_{c}\simeq -l_{loc}\ln (\Gamma l\sqrt{\bar{\varepsilon}}).
\end{equation}
Unlike the lossless case, the disappearance of reflection is caused not by
the high transparency of the sample but rather by strong absorption in the
effective cavity. The distance $x_{c}$ as well as the maximal intensity does
not depend on the total length $L$ of the sample. In this regard, the sample
can be considered as a half-infinite random medium.

Equations (\ref{3a})--(\ref{4}) enable one to formulate an algorithm for
retrieval the internal characteristics of localized states via external
measurements. Indeed, quantities $R_{res}$, $T_{res}$, $\delta \nu $, $%
R_{typ}=1-\Gamma l_{loc}\sqrt{\bar{\varepsilon}}-T_{typ}$, and $T_{typ}=\exp
\left( -L/l_{loc}\right) $ can, in principle, be measured outside the
sample. Then, Eqs. (\ref{3a})--(\ref{4}), constitute a system of equations,
from which two internal parameters of the sample, $\Gamma $ and $l_{loc},$
and three parameters of the resonances, $l,$ $d,$ and $I_{res}$ can be
found. However, even small dissipation can make transmittance non-measurable
(as it was in our experiment), and only reflected signal can be analyzed.
Nonetheless, the developed model still allows determination of parameters of
the sample and resonances. When $b\gg 1$ the sample can be treated as a
half-infinite and one can put $T_{2}=0$ in Eqs.~(\ref{3a})--(\ref{4}) and
obtain the following sample parameters: 
\begin{equation}
\Gamma =\frac{2\pi \delta \nu \sqrt{\bar{\varepsilon}}}{c}\left( 1-\sqrt{%
R_{res}}\right) ,\,\,l_{loc}\simeq \frac{1-R_{typ}}{\Gamma \sqrt{\bar{%
\varepsilon}}}.  \label{9}
\end{equation}%
The absorption rate $\Gamma $ and the localization length $l_{loc}$ are
expressed via readily measured quantities $R_{res}$, $\delta \nu $ and $%
R_{typ}$. The values of $\Gamma $ and $l_{loc}$ calculated by means of Eq.~(%
\ref{9}) with $\delta \nu $ and $R_{res}$ measured for all configurations
are in agreement within $\Gamma =3.76\cdot 10^{-3}cm^{-1}\pm 5\%$ and $%
l_{loc}\simeq 1.26cm$. The central position, $x_{c},$ of a resonance and its
peak intensity, $I_{res},$\ are given by 
\begin{eqnarray}
I_{res} &=&\frac{1-R_{res}}{\Gamma l\sqrt{\bar{\varepsilon}}},  \label{10} \\
x_{c} &=&-l_{loc}\ln \left[ \frac{2\pi \delta \nu l\bar{\varepsilon}}{c}%
\left( 1+\sqrt{R_{res}}\right) \right] .
\end{eqnarray}

To check the algorithm of the retreival of the internal parameters of
resonances from the outside measurements of the reflection coefficient, we
have compared the coordinates, $x_{c}$, and the normalized
intensities, $I_{res}$, of the resonances calculated by Eqs. (\ref{10}%
) (using measured $R_{res}$ and $\delta \nu $) with the
corresponding values measured immediately inside the waveguide by an antenna
incerted into the waveguide through a slot. As examples, the calculated
(first line) and measured (second line) values of $x_{c}$ and $%
I_{res}$ for three resonances depicted in Fig.~1a ($\nu $%
=14.8, 15.5, 16.5 GHz) are given in the following table:

\vspace{2mm}%
\begin{tabular}{l|c|c}
& $x_{c}$(cm) & $I_{res}$ \\ \hline
calc. & 4.0, 4.8, 3.6 & 190, 150, 123 \\ 
meas. & 4.2, 5.5, 3.6 & 86, 201, 96%
\end{tabular}

Note that most of the detected resonances were in vicinity of the critical
coupling regime (white area in Fig.~3b).

To conclude, localized states in a 1D lossy random system has been studied
experimentally and analytically. The theoretical analysis is based on a
model, which describes isolated disorder-induced resonances in the
framework of the general theory of resonant systems. It is shown that when
the dissipation length is much larger than the total length of the sample
the effect of dissipation off resonance is negligible. However, on
resonance, the effect is dramatic and vanishes only when the dissipation
length is exponentially large as compared to the sample length. Otherwise,
the energy at a resonant frequency dissipates substantially due to the
exponentially high intensity of the field inside effective resonant cavity,
resulting in the exponential small resonant transmission. The reflection
coefficient depends non-monotonically on the absorption and drops to zero at
some values of the dissipation length (critical coupling effect). An
algorithm has been developed for retrieval the decrement of absorption and
the localization length, as well as the peak intensities and the positions
of the localized modes inside a sample from the measurements of the
reflection coefficient. Surprisingly, losses improve the ``detectability''
of resonances and change the conditions for their excitation. By changing
the loss one can scan the sample detecting modes excited at different points
of the sample. The algorithm has been tested by comparison to the direct
measurements of the microwave field inside a disordered single-mode
waveguide.

We acknowlege the experimental help of Jerome M. Klosne. The work was
partially supported by NSF grant DMR-0538350.

\end{document}